\documentclass[10pt,twoside]{uz_kgu}
\usepackage{psfrag}

\newcommand{\be}{\begin{equation}}
\newcommand{\ee}{\end{equation}}
\newcommand{\beq}{\begin{eqnarray}}
\newcommand{\eeq}{\end{eqnarray}}

\newcounter{algorithm}[section]

\begin{document}
\renewcommand{\proofname}{ {\hskip\parindent \bf Proof.}}
\renewcommand{\refname}{\small {Literature}}
\renewcommand{\abstractname}{Abstract}
\renewcommand{\figurename}{Fig.}

\def\theequation{\thesection.\arabic{equation}}
\renewcommand{\thesection}{\arabic{section}}
\renewcommand{\thesubsection}{\arabic{section}.\arabic{subsection}}
\renewcommand{\thesubsubsection}{\arabic{section}.\arabic{subsection}.
\arabic{subsubsection}}

\makeatletter
\renewcommand{\@seccntformat}[1]{{\csname the#1\endcsname}.\hspace{0.5em}}
\makeatother

\UDK{}

\ArticleNAME{Black holes and quasiblack holes: Some history and
  remarks
}

\ArticleAUTHOR{Jos\'e P. S. Lemos\\
{\small  Multidisciplinary Center for Astrophysics - CENTRA,
Physics Department,  Instituto Superior T\'ecnico - IST,
Universidade T\'ecnica de Lisboa - UTL, Portugal}}

\ArticleHEAD{Black holes and quasiblack holes \ldots}

\ArticleAUTHORHEAD{Jos\'e P. S. Lemos.}

\makeabstitle

\begin{abstract}
A brief reference to the two Schwarzschild solutions and what Petrov
had to say about them is given.  Comments on how the Schwarzschild
vacuum solution describes a black hole are also provided.  Then we
compare the properties, differences and similarities between black
holes and quasiblack holes.  Black holes are well known. Quasiblack
hole is a new concept.  A quasiblack hole, either nonextremal or
extremal, can be broadly defined as the limiting configuration of a
body when its boundary approaches the body's own gravitational radius
(the quasihorizon).  They are objects that are on the verge of being
black holes but actually are distinct from them in many ways.  We
display some of their properties: there are infinite redshift whole
regions; the curvature invariants remain perfectly regular everywhere,
in the quasiblack hole limit; a free-falling observer finds in his own
frame infinitely large tidal forces in the whole inner region, showing
some form of degeneracy; outer and inner regions become mutually
impenetrable and disjoint, although, in contrast to the usual black
holes, this separation is of a dynamical nature, rather than purely
causal; for external far away observers the spacetime is virtually
indistinguishable from that of extremal black holes.  Other important
properties, such as the mass formula, and the entropy, are also
discussed and compared to the corresponding properties of black holes.

\end{abstract}
\footnotemark{In
the Scientific Proceedings of Kazan State University
(Uchenye Zapiski Kazanskogo Universiteta (UZKGU)) 
{\bf 153}, 351 (2011), ed.~A. 
Aminova. Based on the Invited Lecture 
in the Petrov 2010 Anniversary Symposium on General Relativity and 
Gravitation, Kazan, Russia, November 1-6, 2010.}

\vskip -0.1cm
\vspace{\baselineskip}\hrule

\section{Introduction}\label{S:In}

\subsection{The Schwarzschild solution.}\label{sch}
\hskip0.1cm
Finding vacuum solutions of Einstein's equation
\vskip -0.7cm
\begin{equation}
G_{ab}=0\,,
\end{equation}
\vskip -0.2cm
\noindent 
where $G_{ab}$ is the Einstein
tensor, is an important branch of General Relativity and known to be a
non-trivial task. On the other hand, finding solutions of the field
equations with matter is a somewhat different setup.  Given any
metric, there is always one stress-energy tensor $T_{ab}$ for
which Einstein's equations $(G=1\,,c=1)$
\vskip -0.3cm
\begin{equation}
G_{ab}=8\pi\,T_{ab}\,,
\end{equation}
\vskip -0.2cm
\noindent 
are trivially satisfied.
Now, arbitrarily chosen metrics usually give rise to unphysical
stress-tensors, corresponding to matter which 
is of no interest. 
Therefore, the task of finding non-vacuum solutions to the
field equations is, in a certain way, twice as hard in comparison to
solutions in vacuum, one has to choose physically relevant sources,
and then solve for the gravitational field in the equations.

Schwarzschild, in 1916, in two strokes, 
initiated the field of exact solutions in
General Relativity, both in vacuum \cite{Schw1} and in matter
for an incompressible fluid \cite{Schw2}. 
These solutions are called the Schwarzschild
solution and the interior Schwarzschild solution, respectively.
The
Schwarzschild solution \cite{Schw1} is perhaps the most well-known
exact solution in General Relativity, and its line element can be
written in appropriate spherical coordinates $(t,r,\theta,\phi)$ as,
\begin{equation}
ds^2=-\left(1-\frac{2m}{r}\right)\,dt^2+
\frac{dr^2}{\left(1-\frac{2m}{r}\right)}+
r^2\left(d\theta^2+\sin^2\theta\,d\phi^2\right)\,.
\label{leschw}
\end{equation}
Here $m$ is the mass of the object, outside which there 
is vacuum. To interpret the solution as a whole vacuum
solution, and the emergence of the notion
of a black hole it took some time.

\subsection{Petrov on the Schwarzschild solution.}\label{pet}
\hskip0.2cm
In a Petrov Symposium it is worth to spend some lines on what Petrov
had to say on both Schwarzschild solutions.  
For this we refer to
his book {\it Einstein spaces}, published in Russian 
in 1961 and then translated
into English in 1969 \cite{petrov1}.

On p. 141 of the book \cite{petrov1} one 
can read the rather remarkable phrase: ``It is clear that Einstein,
Hilbert, and their contemporaries had a rather primitive idea of what
is meant by `spacetime metric' and of its scope. They possessed only a
few of the simplest examples (for example Schwarzschild's solution,
the solution of Weyl and Levi-Civita with axial symmetry, and
cosmological metrics). They did not realize what a powerful
instrument they were forging.''

Then there are several mentions, in passing, of the Schwarzschild
solution.  On p. 179 it is stated that the Schwarzschild solution is a
particular case of solutions included in $T_1$, i.e., solutions with
Segre characteristic $(111)$, referring to his algebraic
classification of 1954 of the Riemann and Weyl tensors \cite{petrov2},
repeated in the book in page 99.  On p. 196 Kotler's solution is
mentioned, stating it is a generalization of the Schwarzschild
solution by including a cosmological term $\Lambda$.  On p. 360, in
Chapter 9, Einstein's equations for a spherically symmetric vacuum are
solved, and the Schwarzschild solution is finally displayed, as a
textbook should do.  On p. 362, exercises on Schwarzschild and
interior Schwarzschild are given, and the Landau and Lifshitz 1948
book {\it The Classical Theory of Fields} (and the English translation
of 1959) is cited \cite{landaulifsh}.  On p. 386 the two
Schwarzschild's papers of 1916 on the vacuum and the interior
solutions are quoted in citations 37 and 37a, respectively.

There is an interesting contribution of Petrov to the field of exact
solutions. In the paper {\it Gravitational field geometry as the
geometry of automorphisms}
\cite{petrov3}, among a discussion of many solutions, 
Petrov finds a Type I (111) solution with metric
$
ds^2=e^r\cos\sqrt{3}r(-dt^2+d\phi^2)-2\sin\sqrt{3}r\,d\phi\,dt
+
dr^2+e^{-2r}dz^2
$.
It is the only vacuum solution admitting a simply-transitive
four-dimensional maximal group of motions.  Bonnor \cite{bonnoronpetr}
showed that it is the vacuum solution exterior to an infinite rotating
dust, a particular case of the Lanczos-van Stockum solution.  This is
no black hole but has relations to the hoop conjecture, closed
timelike curves and so on.

\subsection{Black holes.}\label{bhsfinally}
\hskip0.2cm
It is clear that the Schwarzschild solution (\ref{leschw}) presents a
problem, in the coordinates used, at $r=2m$.  For a long time $r=2m$
was a mysterious place. Only in the 1960s the ultimate interpretation
was given and the problem solved. The radius $r_{\rm h}=2m$
defines the event horizon, a lightlike surface, of
the solution. In its full form it represents
a wormhole, with its two phases, the white hole and the black hole,
connecting two asympotically flat universes \cite{kruskal} 
(a work done under the supervision of Wheeler
\cite{wheelbiog}).
If, besides a mass $m$ as in the Schwarzschild solution, 
one includes electrical charge $q$, 
the Reissner-Norstr\"om solution 
is obtained \cite{reiss,nordst}
(for the interpretation of its full form see  \cite{gravesbrill}). 
The inclusion of angular momentum $J$ gives the Kerr
solution \cite{kerr}, and the inclusion of the three parameters
($m,J,q$) is the Kerr-Newman family \cite{newman}. 
For a full account of the Kerr-Newman family
within General Relativity see \cite{mtw}.

As predicted early by Oppenheimer 
and Snyder \cite{oppenheimer} black holes can form 
through  gravitational
collapse of a lump of matter. As the matter falls in, an 
event horizon develops from the center of the matter,
and stays put, as a null surface, in the spherical symmetric case
at $r_{\rm h}=2m$, while the matter falls in towards a
singularity.
A posterior important result 
is that if the matter is made of perfect fluid 
(such as the Schwarzschild interior solution \cite{Schw2})
there is the Buchdahl 
limit \cite{buch} which states that when the boundary of the 
fluid matter approaches quasistatically the value $\frac98\,r_{\rm h}$, 
then the 
system ensues in an Oppenheimer-Snyder collapse, presumably 
into a black hole.

The possibility of existence of black holes came with Quasars in
1963. Salpeter \cite{salpeter} and Zel'dovich \cite{zeldovich} were
the first to advocate that a massive central black hole should be
present in these objects in order to explain the huge amount of energy
liberated by them.  Lynden-Bell in 1969 then took a step forward and
proposed that a central massive black hole should inhabit every galaxy
\cite{lyndenbell}, a prediction that has been essentially confirmed,
almost every galaxy has a central black hole. Then with the discovery
of pulsars in 1968 and the reality of neutron stars the possibility of
small stellar mass black holes became obvious, confirmed in 1973 with
the X-ray binary Cygnus X1 and then with other X-ray binary sources
(see, e.g., \cite{lemosbhsgalelempart}).

It is supposed that black holes can form in many ways.  The
traditional manner is the Oppenheimer-Snyder type collapse
\cite{oppenheimer}.  Nowadays, one also admits that black holes can
form from the collision of particles, or have a cosmological
primordial inbuilt origin (see, e.g., \cite{lemosbhsgalelempart}).
The Reissner-Nordstr\"om black hole 
may be not very useful astrophysically, 
although all black holes should have a tiny, fluctuating,
charge. Notwithstanding it might be important in particle physics, 
perhaps it is an elementary soliton 
of gravitation, as proposed by some supergravity ideas. 
Nowadays there is a profusion of theoretical black 
holes of all types, in all
theories, with all charges, in all dimensions
(see, e.g., \cite{profusion}).

Classically, black holes are 
well understood from the outside: there is astrophysical 
evidence and theoretical consistency.
Perhaps there will be  phenomenological 
evidence in the near future from the collision of
particles.

Quantically, black holes still pose problems.  For the outside, these
problems are related to the Hawking radiation and the
Bekenstein-Hawking entropy.  For the inside, the understanding of the
inside of a black hole is one of the outstanding problems in
gravitational theory, and it certainly is
a quantum phenomenon. The horizon
harbors a singularity. What is a singularity?
The two quantum problems, the outside and the inside,
are perhaps related. There are many approaches, some try
to solve part of the problems others all of them
(see, e.g., \cite{lemosfundamental}). These 
approaches are the quantum gravity approach, mass inflation,
wormhole, regular black hole, holographic reasoning
(see, e.g., \cite{lemosholographic}) and so on. Here, 
we advocate the quasiblack hole approach
to better understand a black hole, both the outside and the inside
stories. We do not claim to solve the problems, we 
look at it through a different angle and see 
where it leads us to.

\subsection{Quasiblack holes.}\label{qbhsfinally}
\hskip0.2cm
Following \cite{buch}, 
for matter made of perfect fluid there is the Buchdahl 
limit. 
However, putting charge into the matter 
to bypass the limit opens up a new world.
The charge can be electrical, or angular momentum, or many other
charges.
The simplest case is to have matter with electric charge alone, 
nothing else.

In Newtonian gravitation, i.e., for a Newton-Coulomb system, 
the solution is easy. Suppose one has two
massive charged particles.
Then, the gravitational force exerted on each 
particle is $F_{\rm g}=\frac{Gm^2}{r^2}$, where 
for a moment we have restored $G$,
and the electric force is $F_{\rm e}=\frac{e^2}{r^2}$.
Thus, when $\sqrt{G}m=e$ it implies $F_{\rm g}=F_{\rm e}$.
The system is in equilibrium.
Of course, if we put another such particle, any number of particles, 
a continuous distribution of matter,
any symmetry, any configuration, the result still holds.
For a continuous distribution the relation 
$\sqrt{G}\rho_{\rm m}=\rho_{\rm e}$ must hold, where 
$\rho_{\rm m}$ and $\rho_{\rm e}$ are the mass-energy density 
and the electric charge density, respectively.

In General Relativity, i.e., for an Einstein-Maxwell
system, the history is long.
Weyl in 1917 \cite{weyl1}
started with a static solution in the form, 
\begin{equation}
ds^2=-W^2(x^i)\,dt^2+g_{ij}(x^k)\,dx^i\,dx^j.
\label{weyl1}
\end{equation}
Then he sought $W$ such that $W^2=W^2(\phi)$,  in vacuum 
with axial symmetry,
where $\phi$ is the electric potential. 
He found $W^2= \left(\,\sqrt{G}\,\phi+b\right)^2+c$,
with $b$ and $c$ constants.
In 1947 Majumdar \cite{maj} showed that 
Weyl's quadratic function works for any symmetry,
not only axial symmetry. 
It was also shown that 
the (vacuum) extremal Reissner-Nordstr\"om solution obeyed this
quadratic relation, and that many such solutions
could be put together since, remarkably, equilibrium 
would be maintained, as in the Newton-Coulomb case. 
Papapetrou \cite{papa} also worked along
the same lines.
Hartle-Hawking in 1973 \cite{hh73} worked out 
the maximal extension
and other properties 
of a number of extremal black holes dispersed in spacetime.
Furthermore,
for a perfect square, 
$W^2(\phi)= \left(\,\sqrt{G}\,\phi+b\right)^2$,
if now there is matter, Majumdar and 
Papapetrou found that  $\sqrt{G}\rho_{\rm m}
=\rho_{\rm e}$ \cite{maj,papa}, and the 
matter is in an equilibrium state, bringing 
into General Relativity the Newtonian result.
This type of matter we 
call extremal charged dust matter.
The solutions, vacuum or matter solutions, 
are generically called Majumdar-Papapetrou solutions.

Now, if one wants to make a star 
one has to put some boundary on the matter.
The interior solution is then Majumdar-Papapetrou
and the exterior is extremal Reissner-Nordstr\"om.
This analysis was started by Bonnor who since 
1953 has called attention to them,
see, e.g., \cite{bonnor1999}.
Examples of Bonnor stars are: (i) A 
star of clouds,
in which each cloud has 1 proton and $10^{18}\,$neutrons,
so to maintain the relation 
$\rho_{\rm m}=\rho_{\rm e}$ ($G=1$). 
For a spherically symmetric star with radius $R$, 
the star as a whole
has $m=q$, and the exterior is extremal Reissner-Nordstr\"om,
see Figure 1.
\begin{figure}[htb]
\label{bonnorstar}
\centering
\includegraphics{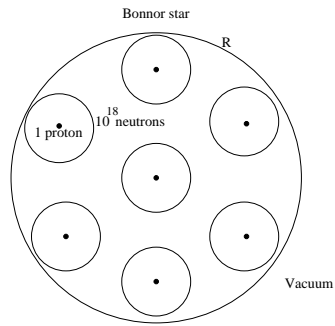}
\caption{A star of clouds as 
an example of a Bonnor star: Majumdar-Papapetrou 
(extremal charged dust) matter
inside, extremal Reissner-Nordstr\"om outside, and a boundary
surface joining the inside and outside at the radius $R$.}
\end{figure}
(ii) A star made of supersymmetric stable particles with $m_{\rm
s}=e_{\rm s}$.  Again, the star has total mass 
$m$ and total charge $q$ related by $m=q$.

Now comes the important point.
For any star radius $R$ the star is in equilibrium. Inclusive
for $R=r_{\rm h}$, where 
$r_{\rm h}=m$ is the gravitational, or horizon, radius of the 
extremal Reissner-Nordstr\"om metric.
What happens when $R$ shrinks to  $r_{\rm h}$?
Something new: a quasiblack hole forms.

\section{Black hole and quasiblack hole solutions}\label{qbhsols}

\subsection{Generic features of the solutions.}\label{bhsqbhssols}
\hskip0.2cm
The difference between an extremal spherically symmetric 
black hole and 
an (extremal) spherically symmetric quasiblack hole
spacetime is best displayed if we write the metric as,
\begin{equation}
ds^{2}=-B(r)\,dt^{2}+A(r)\,dr^{2}+r^{2}\,\left(d\theta
^{2}+\sin ^{2}\theta \,d\phi ^{2}\right)\,.  
\label{metricgeneric}
\end{equation}
When one approaches the gravitational 
radius of the object one finds that the solutions
have the features shown in Figure 2.
\begin{figure}[htb]
\centering
\includegraphics{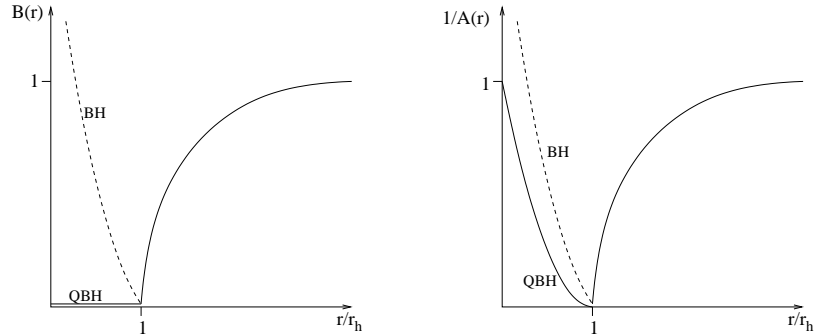}
\caption{Comparison of the generic form 
of the metric potentials $B$ and $A$ 
for black holes and quasiblack holes.}
\end{figure}
For the extremal Reissner-Nordstr\"om black hole one has 
$B(r)=1/A(r)=(1-m/r)^2$, so that at 
$r=r_{\rm h}=m$ there is the usual event horizon, and at $r=0$ 
the potentials are singular and indeed yield a singular
spacetime where the curvature invariants diverge.
For the extremal quasiblack hole the function 
$1/A(r)$ is well behaved, touches zero at $r=r_{\rm h}$,
when a quasihorizon (not an event horizon) forms, 
and tends to 1 at $r=0$ so that there are no conical 
singularities. The function $B(r)$ is well-behaved up
to the quasiblack hole limit.
At the quasiblack hole limit, $R=r_{\rm h}$, the function
is zero 
in the whole interior region. This brings new features.

\subsection{Black holes and quasiblack holes
made of Majumdar-Papapetrou stuff.}\label{lemosweinberg}
\hskip0.2cm
The Majumdar-Papapetrou vacuum black hole is the extremal 
Reissner-Nordstr\"om black hole, a solution 
with well known properties.

For quasiblack holes, Majumdar-Papapetrou matter
provides perhaps the simplest case,  as shown
by 
Lemos and Weinberg in 2004 \cite{lemosweinberg2004}.
In \cite{lemosweinberg2004} it was found a solution in which 
there is no need for a junction. In the 
solution, the Majumdar-Papapetrou
matter decays sufficiently rapid to yield at infinity,
in a continuous way, the extremal 
Reissner-Nordstr\"om metric. In this way the existence
of simple quasiblack hole solutions were shown beyond
doubt. The potentials and all their derivatives are 
continuous. Thus one avoids the possible problems caused 
by Bonnor stars where the potentials are only
one time differentiable. 
To find the 
solutions, insist on putting the metric as in 
Eq.~(\ref{metricgeneric}).
Then Einstein-Maxwell equations give
\begin{equation}
\frac{(AB)'}{AB}=8\pi\,r\,\rho\,A
    \,,\quad \left[r\left(1-\frac{1}{A}\right)\right]'
     =8\pi\,r^2\,\rho+\frac{r^2}{A\,B}\,{\varphi'}^2\,,
\label{equationforB-EM}
\end{equation}
\begin{equation}
    \frac{\sqrt{B}}{r^2\sqrt{AB}}\left[
    \frac{r^2}{\sqrt{AB}}\,\varphi'\right]'
    = - 4\pi\rho_{\rm e}
    \,.
\label{equationfophiEM}
\end{equation}
where primes denote differentiation with respect to $r$.
One can then work out the various types of solutions 
\cite{lemosweinberg2004}. 
\begin{figure}[htb]
\includegraphics[height=2.2cm]{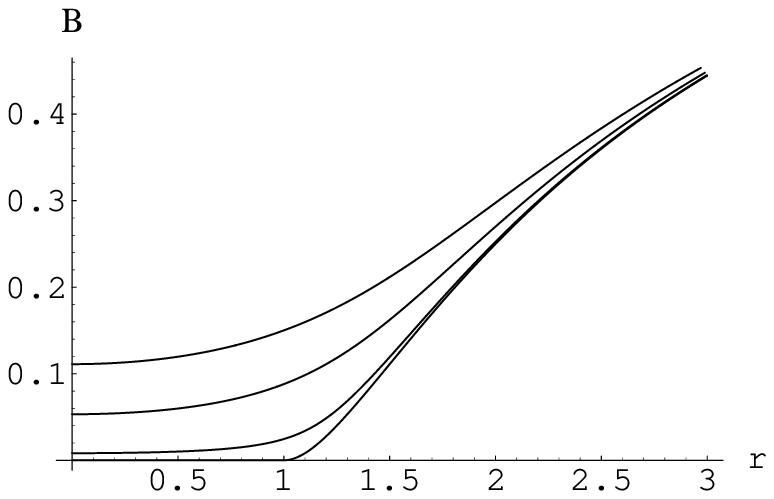}
\includegraphics[height=2.2cm]{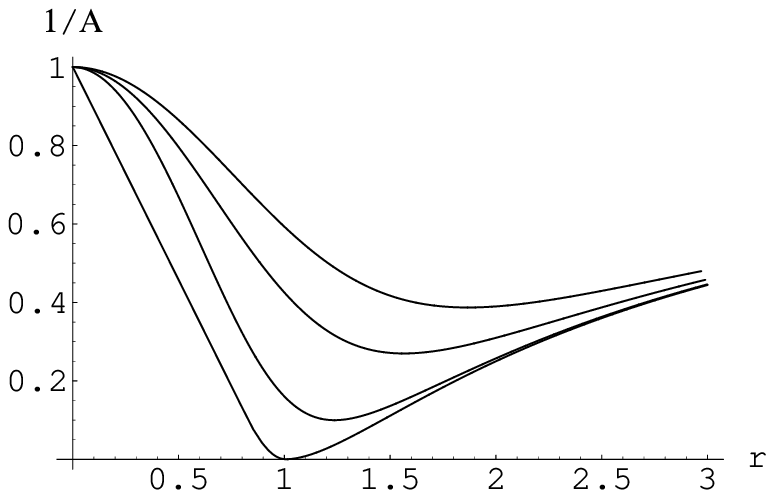}
\includegraphics[height=2.2cm]{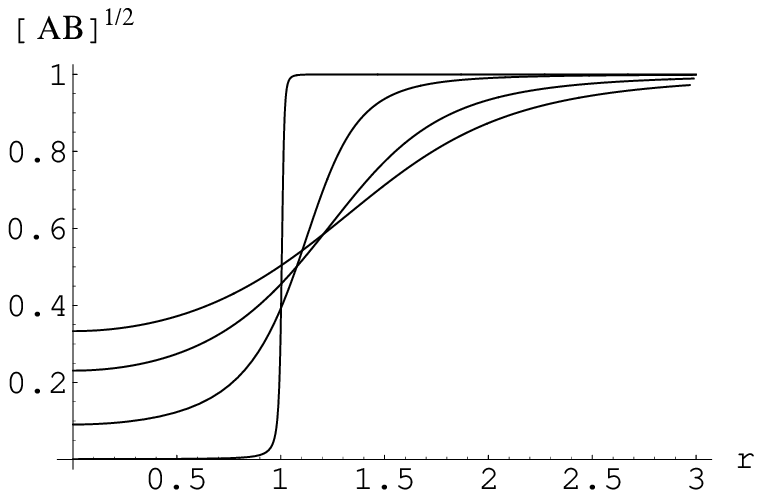}

\includegraphics[height=2.2cm]{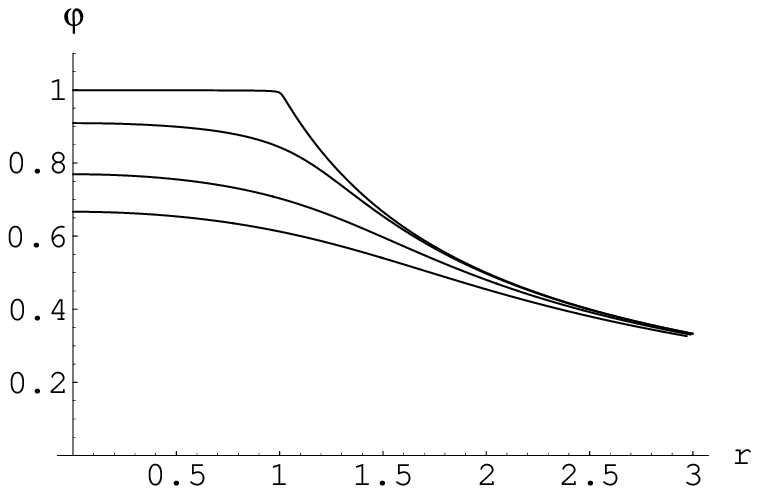}
\includegraphics[height=2.2cm]{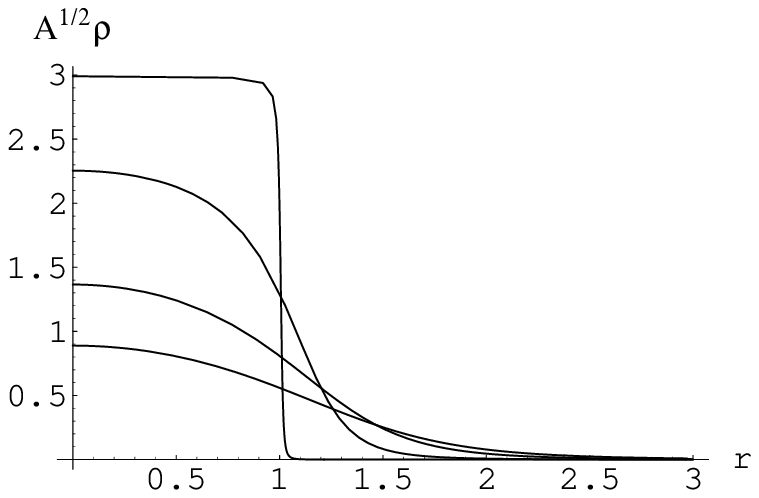}
\caption{Plots of the potentials and matter functions as a function of
$r$ for $q=1$ and for four different stars, each with compact
parameter $c$ given by $c=0.5,\,0.3,\,0.1,\,0.001$.  The
emergence of the quasihorizon is quite evident in the
$c=0.001$ curve, see \cite{lemosweinberg2004} for details.}
\end{figure}
These stars have no well-defined
radius $R$ since there is no boundary. The solutions
tend smoothly into the extremal Reissner-Nordstr\"om 
vacuum. Instead, there is
a compact parameter
$c$ which characterizes the solution. As this 
parameter tends to zero, $c\to0$, the star
gets denser at the center and more compact. 
At $c=0$ a quasiblack hole appears. This is shown 
in Figure 3, where plots for four different stars
(i.e., with stars with different $c${\small s}) are 
displayed.  The one in which $c\to0$ 
shows clearly a quasiblack hole behavior, with 
the emergence of a quasihorizon.

\subsection{Other ways: Black holes and quasiblack holes made 
of various sorts of matter.}\label{otherways}
\hskip0.2cm 
There are black hole solutions in general relativity other than the
ones provided by the Kerr-Newman family.  Those are regular black
holes in which the vacuum inside the horizon with its singularity is
replaced by a de Sitter core, which can be magnetically charged
\cite{bardeen1968} or have non-isotropic pressures \cite{dymn1992}, or
have some other form (see, e.g., \cite{ansoldi2008}).  
There are also regular
black holes electrically charged in a special way
\cite{lemoszanchin2011}.
Black holes are generic. 

What about quasiblack holes? 
Can they be built from other configurations and  forms 
of matter other than Majumdar-Papapetrou? Yes, there 
are several different quasiblack hole solutions found 
up to now.

First, there are the simple quasiblack holes of Lemos and Weinberg,
already mentioned \cite{lemosweinberg2004}.

Second, spherical Bonnor stars (charged stars with a  
spherical boundary surface)
also yield quasiblack holes. This was shown preliminary 
by Bonnor 
himself (see, e.g., \cite{bonnor 1999}) and by subsequent 
works \cite{kleberlemoszanchin2005,lemoszanchin2008}.
Moreover, recently Bonnor has shown
that spheroidal stars made of extremal charged dust 
tend in the appropriate limit 
to quasiblack holes \cite{bonnor2010}.
Generic properties of the Majumdar-Papapetrou matter
in d-dimensions were displayed in 
\cite{lemoszanchin2005}.

Third, charged matter with pressure 
(with a generalized Schwarzschild interior ansatz
to include electrical charge)
also yield 
charged stars that when sufficiently compact tend
to quasiblack holes. These are the relativistic 
charged spheres which can then be considered
as the frozen stars
\cite{lemoszanchin2010,felice1995,felice1999}. For the 
properties of the solutions  and the connection
with the Weyl-Guilfoyle ansatz \cite{guilf1999}
see \cite{lemoszanchin2009prop}. These 
solutions have additional interest
since the pressure stabilizes the fluid
against kinetic perturbations.

Fourth, the Einstein---Yang-Mills--Higgs equation yield 
gravitationally magnetic monopoles that when 
sufficiently compact, form, in certain instances, 
quasiblack holes, as shown by 
Lue and Weinberg 
\cite{lueweinberg1999,lueweinberg2000}. In these 
works the name quasiblack hole was coined
for the first time. A comparison between 
gravitationally magnetic monopole and 
Bonnor star behavior was done in \cite{lemoszanchin2006}.

Fifth, the Einstein-Cartan system with spin and torsion, 
in which 
the spinning matter, put in a spherically symmetric 
configuration, is joined into the Schwarzschild 
solution, also yields quasiblack holes
\cite{lemoseinsteincartanqbh2011}.

Sixth, disk matter systems, when sufficiently compact 
and rotating at the extremal 
limit have, as exterior metric, the extremal Kerr spacetime.
These solutions were found by Bardeen and Wagoner 
back in 1971 \cite{bardeenwagoner1971}. 
In the new language they are quasiblack holes
and their properties have been explored by
Meinel and collaborators \cite{meinel2006,meinel2010}.

Finally, it is a simple exercise to show that 
a shell of matter, for which  the inside is a 
Minkowski spacetime, and the outside  is Schwarzschild, 
yields solutions with quasiblack hole
properties if the shell is allowed to hover 
on the quasihorizon. A drawback here, that does not 
appear in the six mentioned cases above, 
is that in the quasihorizon limit the tangential
pressures grow unbound. We will comment on this 
when we work out the mass formula for quasiblack holes.

There are certainly many other examples in which 
quasiblack holes may form.

\section{Black holes and quasiblack holes: 
Definition and properties}\label{defsprops}

\subsection{Black holes.}\label{bhsdef}
\hskip0.2cm
Black hole definition can be seen in 
\cite{pen72,hawkhouches,hawkingellis}. 
Some of the black hole properties were developed in, e.g., 
\cite{cart1979,bardcarhawk73,smarr1973,beken1973,hawktemp,by1993}.

\subsection{Quasiblack holes.}\label{defiprop}
\hskip0.2cm Since it appears that quasiblack hole solutions
are more ubiquitous than one could 
have thought, one should consider
the core properties of those solutions the most independently
as possible from the matter they are made, in much the same 
way as one does for black holes
\cite{lzs07,lzs08mim,lzs10p,lzs08m1,lzs09m2,lzs10e1,lzs10e2}.

\subsubsection{Definition.} 
\hskip0.2cm
Write the metric as in Eq.~(\ref{metricgeneric}), 
for an interior metric with an asymptotic flat exterior region.
Consider the solution satisfies the following requisites:
(a) the function $1/A(r)$ attains a minimum at some
$r^{\ast }\neq 0$, such that $1/a(r^{\ast })=\varepsilon$, with $\varepsilon
<<1$.  
(b) For such a small but nonzero $\varepsilon$
the configuration is regular everywhere with a nonvanishing metric function
$B$. 
(c) In
the limit $\varepsilon \rightarrow 0$ the metric coefficient $B\rightarrow
0$ for all $r\leq r^{\ast}$. See Figure 2.
These three features define a quasiblack hole \cite{lzs07}.
The quasiblack hole is on the verge of forming an event horizon, but 
instead, a quasihorizon appears with $r^{\ast}=r_{\rm h}$. The metric is 
well defined everywhere and the spacetime should be regular
everywhere. One can try to give an 
invariant definition of a quasiblack hole instead. For instance,
in (a) one can replace $1/A$ by $(\nabla r)^{2}$.
Note that this definition shows that the quasihorizon 
is related to an apparent horizon 
\cite{hawkhouches}
rather than to an event horizon.

\subsubsection{Generic properties.} 
\hskip0.2cm
A study of the several 
properties that can be deduced 
from the above definition was initiated by 
Lemos and Zaslavskii \cite{lzs07}.
Some generic properties are:
(i) The quasiblack hole is on the verge of forming an event horizon, instead, 
a quasihorizon appears.
(ii) The curvature invariants remain regular everywhere.
(iii) A free-falling observer finds in his frame infinite
tidal forces at the interface 
showing some form of degeneracy. 
The inner region is, in a sense, a mimicker of a singularity.
(iv) Outer and inner regions become somehow mutually impenetrable and
disjoint. E.g., in the Lemos-Weinberg solution
\cite{lemosweinberg2004}, the interior is
Bertotti-Robinson, the 
quasihorizon region is extremal Bertotti-Robinson,
and the exterior is extremal RN \cite{lzs07}.
(v) There are infinite redshift whole 3-regions.
(vi) For far away observers the spacetime is 
indistinguishable from that of black holes.
(vii) Quasiblack holes with finite stresses must be extremal to the
outside. 

A comparison of quasiblack holes with other objects,
such as wormholes, that 
can mimick black hole behavior was given in 
\cite{lzs08mim}.

\subsubsection{Pressure properties.} 
\hskip0.2cm
One can also work out what conditions 
the matter pressure should obey at the 
boundary when the configuration approaches the 
quasiblack hole regime. For these interesting properties
see \cite{lzs10p}.

\subsubsection{The mass formula.} 
\hskip0.2cm
To find the mass of a quasiblack hole
one develops the Tolman formula 
$m=\int (-T_{0}^{0}+T_{i}^{i})\sqrt{-g}\,d^{3}x$,
where $i$ stands for spacelike indices $1,2,3$. Since 
one uses the energy-momentum tensor $T_{ab}$ of the matter,
this formula is not applicable for vacuum black holes,
for black holes one has to use other methods
\cite{bardcarhawk73,smarr1973}.
Nevertheless, in the  general stationary case, we obtain 
in the horizon limit \cite{lzs08m1,lzs09m2}
\begin{equation}
m=\frac{\kappa A}{4\pi}+2\omega_{\mathrm{h}}J+
\varphi_{\mathrm{h}}q\,,
\end{equation}
where $\kappa$ is the surface gravity, $A$ is the horizon 
area, $\omega_{\mathrm{h}}$ is the horizon angular velocity,
$J$ the quasiblack hole angular momentum, 
$\varphi_{\mathrm{h}}$ the electric potential, and $q$
the quasiblack hole electrical charge.
This is precisely 
Smarr's formula \cite{smarr1973}, but now for
quasiblack holes. 
The contribution of the term $\frac{\kappa A}{4\pi}$
comes from the tangential pressures that grow unbound
at the quasiblack hole limit 
but are at the same time redshifted away to give 
precisely $\frac{\kappa A}{4\pi}$.
For the extremal case, the term $\frac{\kappa
A}{4\pi}$ goes to zero, since $\kappa$ is zero.
See also Meinel \cite{meinel2006,meinel2010}
for the pure stationary solution 
of the Bardeen-Wagoner type disks \cite{bardeenwagoner1971}.

\subsubsection{The entropy.} 
\hskip 0.2cm
To find the entropy 
one uses the first law of thermodynamics 
together with the Brown-York formalism 
\cite{by1993}. The approach developed is 
model-independent, it solely explores
the fact that the boundary of the 
matter almost coincides with quasihorizon
\cite{lzs10e1,lzs10e2}.

For nonextremal quasiblack holes, when one carefully takes 
the horizon limit, 
one finds that the entropy $S$ is
\cite{lzs10e1}
\begin{equation}
S=\frac14 A\,, 
\end{equation}
where $A$ is the quasihorizon area, 
in accord with the black hole 
entropy \cite{beken1973,hawktemp}. The contribution to this value
comes again from the tangential stresses that grow unbound in the
nonextremal case. Since these divergent stresses are at the boundary,
the result suggest that the degrees of freedom are on the horizon.  It
is precisely when a quasihorizon is achieved and the system has to
settle to the Hawking temperature that the entropy has the value
$A/4$.  The result, together with the approach, suggest further that
the degrees of freedom are ultimately gravitational modes. Since the
tangential pressures grow unbound here, all modes, presumably quantum
modes, are excited. In pure vacuum, as for a simple black hole, they
should be gravitational modes.

For extremal quasiblack holes the stresses are finite at the
quasihorizon. So one should deduce that not all possible modes are
excited. This means that the entropy of an extremal quasiblack hole,
and by continuity of an extremal black hole, should be $S\leq\frac14
A$. Indeed in \cite{lzs10e2} we find for extremal
quasiblack holes,
\begin{equation}
0<S\leq\frac14
A\,.
\end{equation}
The problem of entropy for extremal black holes is a particularly
interesting one. Arguments based on periodicity of the Euclidean
section of the black hole lead one to assign zero entropy in the
extremal case. However, extremal black hole solutions in string theory
typically have the conventional value given by the the
Bekenstein-Hawking area formula $S=A/4$. We find an 
interesting compromise.

\section{Conclusions}

Black holes are generic and stable. Quasiblack holes perhaps not. Any
perturbation would lead them into a black hole, although the inclusion
of pressure may stabilize the system.

However, stable or not, the quasiblack hole approach
can elucidate many features of black holes such as 
the mass formula and the entropy.
The quasiblack hole approach to the understanding of black hole
physics seems somehow like the membrane paradigm
\cite{thorne}.  Indeed, by taking
a timelike matter surface into a null horizon, in a limiting process,
we are recovering the membrane paradigm.  One big difference is that
our membrane is not fictitious like the membrane of the membrane
paradigm, it is a real matter membrane.

\vspace{\baselineskip}

\noindent {\bf Acknowledgments.}  
I would like to thank Vilson Zanchin (S\~ao Paulo) and 
Oleg Zas\-lavskii (Kharkov) for the many works in collaboration
related to quasiblack holes. 
I thank Alexander Balakin and Asya Aminova for inviting me
to the Petrov Anniversary Symposium in Kazan held in November 2010.
I thank the interest in my talk by Gennady Bisnovatyi-Kogan 
who raised the important point connected with the
stability of quasiblack holes and suggested a way of 
accessing the problem. I also 
thank the interest in my talk by
Mikhail Katanaev and Dieter Brill
who raised the important point connected with the
Penrose diagrams of quasiblack holes. 
One of my motivations to come 
to this Petrov Anniversary Symposium was a work on 
the Petrov classification of tensors with four indices, such 
as the Levi-Civita tensor, with my student Andr\'e Moita.
For personal reasons, 
he was not able to participate in the Symposium, so 
our contribution has not appeared in it. Nevertheless, we
will publish the results elsewhere. 
I also thank for financial support the Funda\c{c}\~{a}o 
para a Ci\^{e}ncia e Tecnologia (FCT)
through projects
CERN/FP/109276/2009 and PTDC/FIS/098962/2008 and the 
grant SFRH/BSAB/987/2010, 
and the Reitoria da Universidade T\'ecnica de Lisboa 
for specific support to the presentation of my 
talk at the Petrov Symposium.

\renewcommand{\refname}{\small {Literature}}

\end{document}